\documentclass[12pt]{iopart}
\usepackage{graphicx}
\usepackage{subcaption}
\usepackage{siunitx}
\usepackage{todonotes}
\usepackage{svg}
\usepackage{cite}
\usepackage{hyperref}
\usepackage{color}
\usepackage{array}
\usepackage{amssymb}

\newcolumntype{L}{>{$}l<{$}}

\begin{document}
\title[CARS measurements in nitrogen]{Vibrational CARS measurements in a near-atmospheric 
pressure plasma jet in nitrogen: II. Analysis}

\author{J Kuhfeld, D Luggenhölscher, U Czarnetzki}

\address{Ruhr University Bochum, Institute for Plasma and Atomic Physics, Germany}

\ead{jan.kuhfeld@rub.de}

\begin{abstract}
	The understanding of the ro-vibrational dynamics in molecular (near)-atmospheric pressure
	plasmas is essential to investigate the influence of vibrational excited molecules on the 
	discharge properties.
	In a companion paper \cite{kuhfeld_vibrational_nodate}, results of ro-vibrational coherent anti-Stokes
	Raman scattering (CARS) measurements for a nanosecond pulsed plasma jet consisting of two 
	conducting molybdenum electrodes with a gap of \SI{1}{mm} in nitrogen at \SI{200}{mbar}
	are presented.
	Here, those results  are discussed and compared to theoretical predictions based on rate coefficients 
	for the relevant processes found in the literature. 
	It is found, that during the discharge the measured vibrational excitation agrees well with
	predictions obtained from the rates for resonant electron collisions calculated by Laporta \etal \cite{laporta_electron-impact_2014}.
	The predictions are based on the electric field during the discharge, measured by EFISH \cite{kuhfeld_vibrational_nodate,lepikhin_electric_2020}
	and the electron density which is deduced from the field and mobility data calculated with Bolsig+ \cite{hagelaar_solving_2005}.
	In the afterglow a simple kinetic simulation for the vibrational subsystem of 
	nitrogen is performed and it is found, that the populations of vibrational excited states develop
	according to vibrational-vibrational transfer on timescales of a few \si{\micro\second},
	while the development on timescales of some hundred \si{\micro\second} is determined by 
	the losses at the walls. No significant influence of electronically excited states on the 
	populations of the vibrational states visible in the CARS measurements ($v \lesssim 7$) was observed.
\end{abstract}


\submitto{\JPD}

\maketitle

\section{Introduction}
\label{sec:intro}
In recent years, (near-) atmospheric pressure plasmas have gained a lot of interest
in the plasma community due to their numerous possible applications. Those plasmas
often employ relatively complex gas mixtures including molecular gases. Depending 
on the discharge conditions a significant amount of the energy input can be 
stored in the vibrational excitation of molecules, which can potentially enhance 
chemical reactions leading to possible use cases in plasma chemistry \cite{fridman_plasma_2008} and plasma 
catalysis \cite{neyts_understanding_2014,whitehead_plasmacatalysis_2016}. To investigate the influence of the vibrational excitation it is critical
to know the vibrational distribution function. Besides tunable diode laser absorption
spectroscopy (TDLAS), Fourier transform infrared (FTIR) spectroscopy and 
spontaneous Raman scattering, coherent anti-Stokes Raman scattering (CARS) is 
a popular technique to measure the vibrational excitation in gaseous media which
was already employed in plasmas in the past \cite{lempert_coherent_2014}. Nonetheless, CARS has a drawback in 
the sense that the measured signal does depend on the population differences between 
vibrational states, i.e. the populations cannot be determined directly from the 
measured CARS spectra. This is not a major problem only for equilibrium systems where 
the vibrational population densities follow a Boltzmann distribution. With the 
knowledge of the distribution function theoretical CARS spectra can be calculated 
and fitted to measured ones with the temperature as fitting parameter \cite{eckbreth_laser_1996}. The use 
of a simple Boltzmann distribution function is not possible in non-equilibrium 
systems as low-temperature plasmas. Here, several different evaluation methods were
used in the past. One approach is to estimate the population density of one state 
and use this as starting point to calculate the others with the population differences
obtained from the CARS spectra. This is done for example in an early work related 
to CARS in plasmas by Shaub \etal \cite{shaub_direct_1977} where the densities of the 
first two vibrational states were estimated by assuming a Boltzmann distribution for those. 
In other works\cite{deviatov_investigation_1986,montello_picosecond_2013} 
it is assumed that the upper state for the highest detectable transition is approximately
zero. The latter approximation is certainly reasonable for vibrational temperatures
close to room temperature where the ratio between two neighboring states is about
\mbox{$\frac{N_{v+1}}{N_v} \approx \exp\left(-\frac{\hbar\omega_e}{k_B T_{vib}}\right)\approx 1.2\times 10^{-5}$ }
with \mbox{$\omega_e\approx \SI{2359}{cm^{-1}}$} and \mbox{$T_{vib}=\SI{300}{K}$}. This changes drastically for 
higher temperatures, e.g. \mbox{$T_{vib}=\SI{5000}{K}$} where $\frac{N_{v+1}}{N_v}\approx 0.5$. 
For this reason in our companion paper \cite{kuhfeld_vibrational_nodate} a different approach is used.
Similar to CARS measurements in thermal equilibrium a distribution function is assumed for 
fitting theoretical spectra to measured ones, like it was done already by Messina \etal \cite{messina_study_2007}
in a plasma burner. They assume that both the vibrational and the rotational decree 
of freedom are Boltzmann distributed, but have different temperatures. This works 
reasonably well for their measurements as they measure only up to the third vibrational 
state to the limited spectral range. In our measurements in \cite{kuhfeld_vibrational_nodate} and 
previous CARS measurements in different plasma sources it can be seen that usually 
the higher vibrational states are overpopulated compared to a Boltzmann distribution 
determined by the first two or three vibrational states. For this reason a distribution 
function is chosen which includes two vibrational temperatures and one rotational, 
either called vibrational two-temperature distribution or simply two-temperature 
distribution function (TTDF) in the following. As the TTDF is motivated by the
underlying plasma physics in the discharge, a detailed derivation was omitted in \cite{kuhfeld_vibrational_nodate}
where the focus was on the diagnostic method. In the present paper we 
motivate the use of the TTDF based on the excitation processes in the plasma in 
section \ref{sec:discharge_phase}. Additionally, simple models are derived for 
the parameters in the TTDF connecting them to the plasma parameters.
Finally, the population difference results from \cite{kuhfeld_vibrational_nodate} in the afterglow where the TTDF 
is not valid anymore are compared to a simple kinetic model for the vibrational 
subsystem in section \ref{sec:afterglow}.

\section{Summary of the results in \cite{kuhfeld_vibrational_nodate}}
\label{sec:summary}
\begin{figure}
	\centering
	\includegraphics[width=\linewidth]{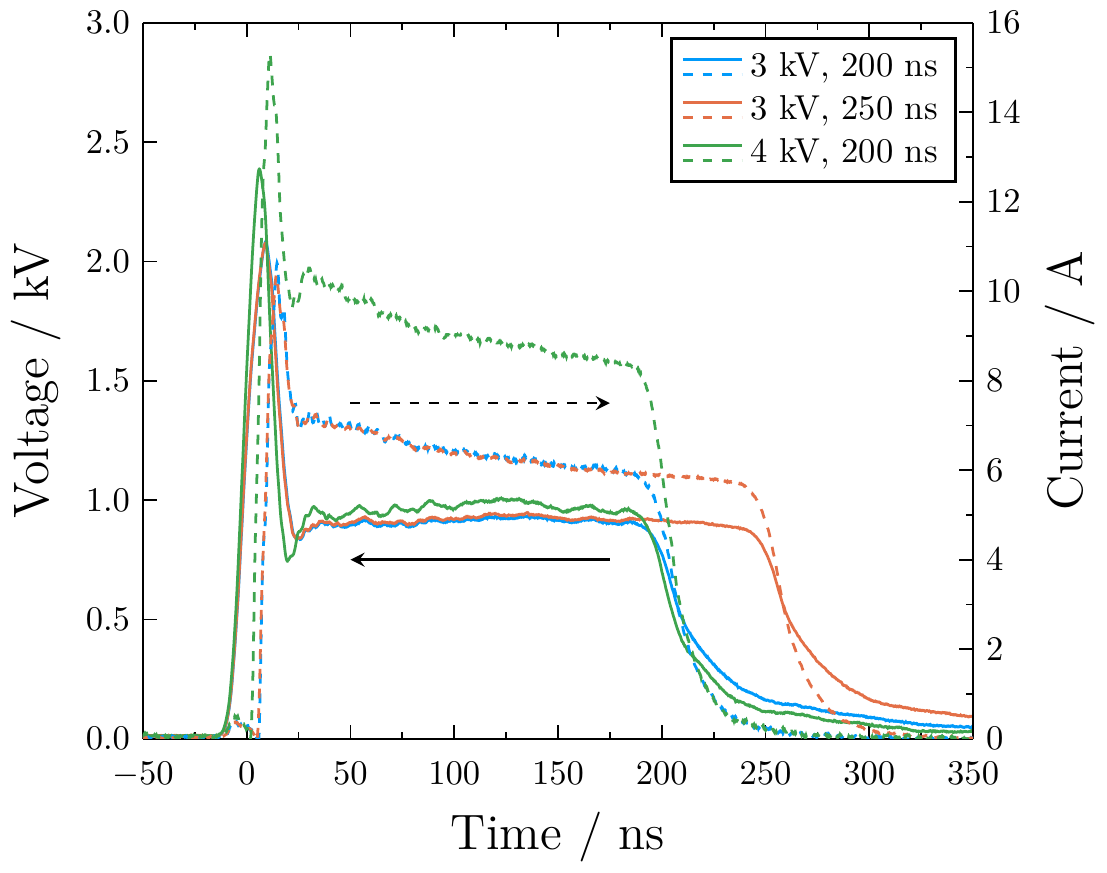}
	\caption{Voltages (solid) and currents (dashed) at the cathode used for the measurements 
	in \cite{kuhfeld_vibrational_nodate}.}
	\label{fig:VI_waveforms}
\end{figure}
In \cite{kuhfeld_vibrational_nodate} a ns-pulsed discharge with a parallel electrode 
configuration is studied. The discharge jet consists of two molybdenum electrodes
with a length of \SI{20}{mm} and a thickness of \SI{1}{mm}. The \SI{1}{mm} gap 
between the electrodes is enclosed with two glass plates at the long edges of the
electrodes. In the middle of one glass plate a hole with about \SI{1}{mm} diameter
serves as gas inlet. This way two opposing gas channels with a
$\SI{1}{mm}\times\SI{1}{mm}$ cross section and \SI{10}{mm} length are constructed.
By applying a high voltage to one of the electrodes while grounding the other 
an electric field can be created transverse to the gas flow. This geometry 
was chosen to provide a simple model geometry for nanosecond pulsed discharges in 
a pressure range close to atmospheric pressure (about \SI{100}{mbar} up to 
\SI{1}{bar}). It should be noted that in contrast to the dielectric barrier discharges (DBD)
more often used to operate nanosecond pulsed plasmas around atmospheric pressure,
here a conduction current can flow for as long as several hundred nanoseconds depending on the power supply.
This was seen to have major impact on the vibrational excitation in \cite{kuhfeld_vibrational_nodate}.\\
The discharge is operated in pure nitrogen at a pressure of  \SI{200}{mbar} and 
a gas flow of \SI{20}{sccm}. A high voltage pulse is applied with \SI{200}{ns} or
\SI{250}{ns} pulse length (see figure \ref{fig:VI_waveforms}) and 
a repetition rate of \SI{1}{kHz}. The applied voltages lead to the current waveforms
depicted in figure \ref{fig:VI_waveforms} with a high current peak during the
ignition of the discharge and a relatively constant, lower current plateau for the 
remainder of the discharge pulse due to a voltage drop at a series resistor. 
These condition produce a homogeneous discharge along the gas channel without arcing. 
The electric field is measured in the middle of the discharge at multiple times
during the constant current plateau via the E-FISH technique and measurements of 
the ro-vibrational distributions are performed by coherent anti-Stokes Raman scattering
(CARS) during the discharge pulse and in the afterglow between two pulses for three 
different voltage pulses shown in figure \ref{fig:VI_waveforms}. 
The electric field is found to be around \SI{81}{Td} at all times during the 
current plateau for all three discharge conditions. 
For the analysis of the measured CARS spectra a vibrational two-temperature distribution
of the form 
\begin{eqnarray}
	N(v,J; T_{rot}, T_{vib,c}, T_{vib,h}, R_{h}) = 
	g(J) \mathrm{e}^{-\frac{E(v,J) - E(v,0)}{k_BT_{rot}}} \\
	\times\Bigl[
		\frac{1-R_h}{Z_c}\mathrm{e}^{-\frac{E(v,0)-E(0,0)}{k_BT_{vib,c}}} 
		+ \underbrace{\frac{R_h}{Z_h}\mathrm{e}^{-\frac{E(v,0)-E(0,0)}{k_BT_{vib,h}}}}_{\text{for }v>0}
	\Bigr] \nonumber
	\label{eq:N_twotemp}
\end{eqnarray}
is proposed which distinguishes between the \mbox{rotational(-translational)} temperature 
$T_{rot}$ and two vibrational temperatures - $T_{vib,c}$ for a vibrationally cold 
distribution and $T_{vib,h}$ for a smaller, vibrationally hot distribution.
For nitrogen the degeneracy of the rotational states is 
\begin{equation}
	g(J) = \left\{\begin{array}{@{\kern2.5pt}lL}
			\hfill 6(2J+1),& \text{if } J \text{ even} \\
			\hfill 3(2J+1),&  \text{if } J \text{ odd.}
		\end{array}\right.
\end{equation}
The partition function for the vib. cold molecules is given by
\begin{equation}
	Z_c = \sum_{v,J} g(J) \mathrm{e}^{-\frac{E(v,J) - E(v,0)}{k_BT_{rot}}}
	\times\mathrm{e}^{-\frac{E(v,0)-E(0,0)}{k_BT_{vib,c}}}
	\label{eq:Z_c}
\end{equation}
and for the vib. hot molecules by
\begin{equation}
	Z_h = \sum_{v>0,J} g(J) \mathrm{e}^{-\frac{E(v,J) - E(v,0)}{k_BT_{rot}}}
	\times\mathrm{e}^{-\frac{E(v,0)-E(0,0)}{k_BT_{vib,h}}}.
	\label{eq:Z_h}
\end{equation}
A detailed motivation of this distribution function is given in section \ref{sec:discharge_phase}
where also the measurement results are presented in figures \ref{fig:Rh_discharge}
and \ref{fig:Texc_fit} together with predictions from 
data for vibrational excitation by resonant electron collisions. It should be noted,
that here and in the following particle densities are always normalized to one, i.e.
they are divided by the gas density.\\
As \eref{eq:N_twotemp} is motivated by the excitation processes in the discharge 
it is not adequate to describe the afterglow. Therefore, there only the population 
differences $\Delta N_{v} = N_{v} - N_{v+1}$ are inferred from the CARS spectra
and the individual population densities are obtained by extrapolating the 
number density of the upper state of the highest detectable transition.
For more details see \cite{kuhfeld_vibrational_nodate}.
Some results are shown in section \ref{sec:afterglow} where they are compared
with a simple volume averaged model for the vibrational system.

\section{Description of the vibrational dynamics}
\label{sec:vib_dynamics}

\subsection{Discharge phase}
\label{sec:discharge_phase}

\begin{figure}
	\centering
	\includegraphics[width=\linewidth]{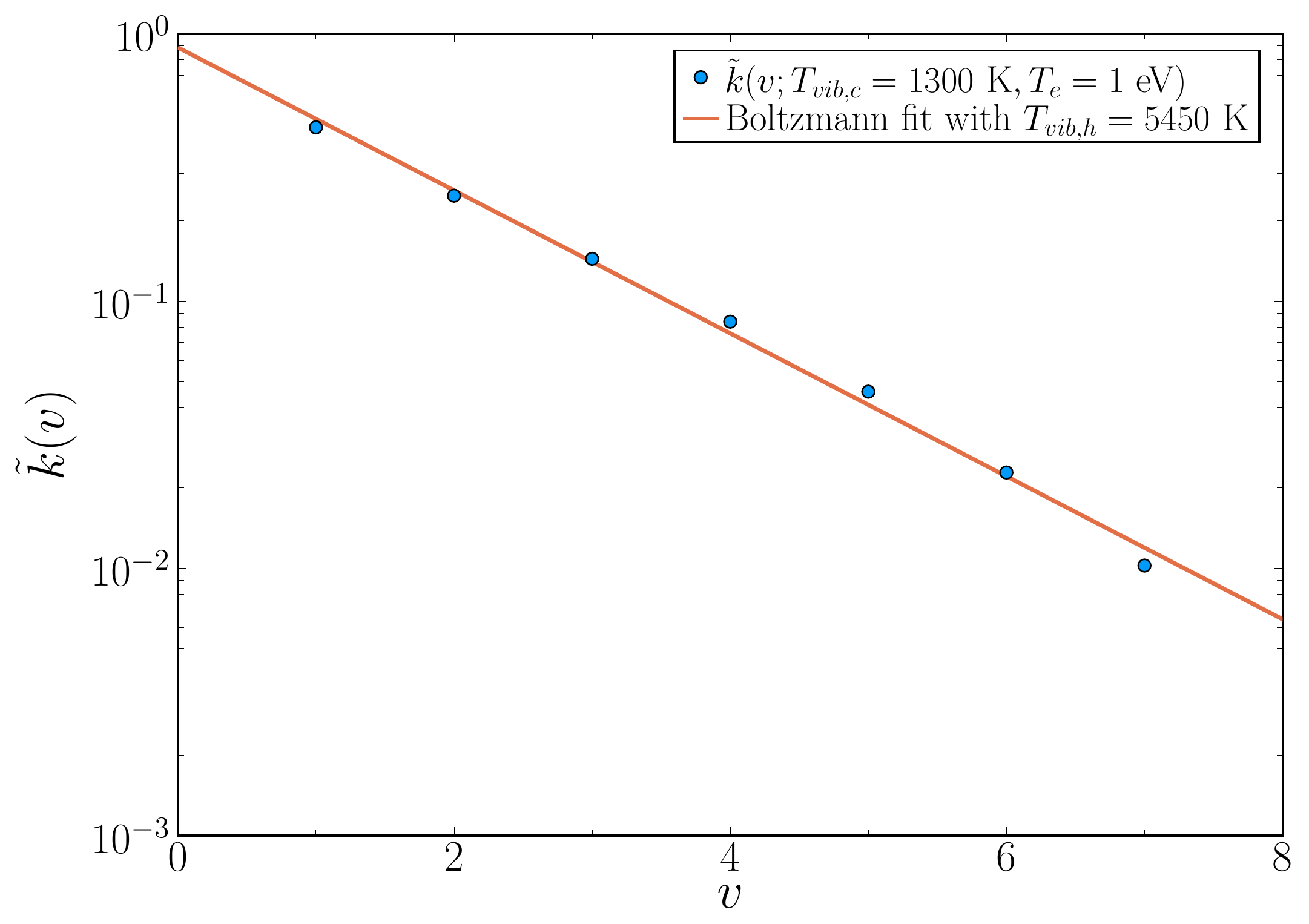}
	\caption{Sum of excitation rates into vibrational states $v$ for an electron
	temperature of $T_e=\SI{1}{eV}$ and a vibrational background with a 
	(vib.) temperature of \SI{1300}{K} (normalized by $\sum_v \tilde{k}(v)$). The solid line is a Boltzmann distribution
	with a temperature of \SI{5450}{K}.}
	\label{fig:Texc_fit}
\end{figure}

\begin{figure}
	\centering
	\includegraphics[width=\linewidth]{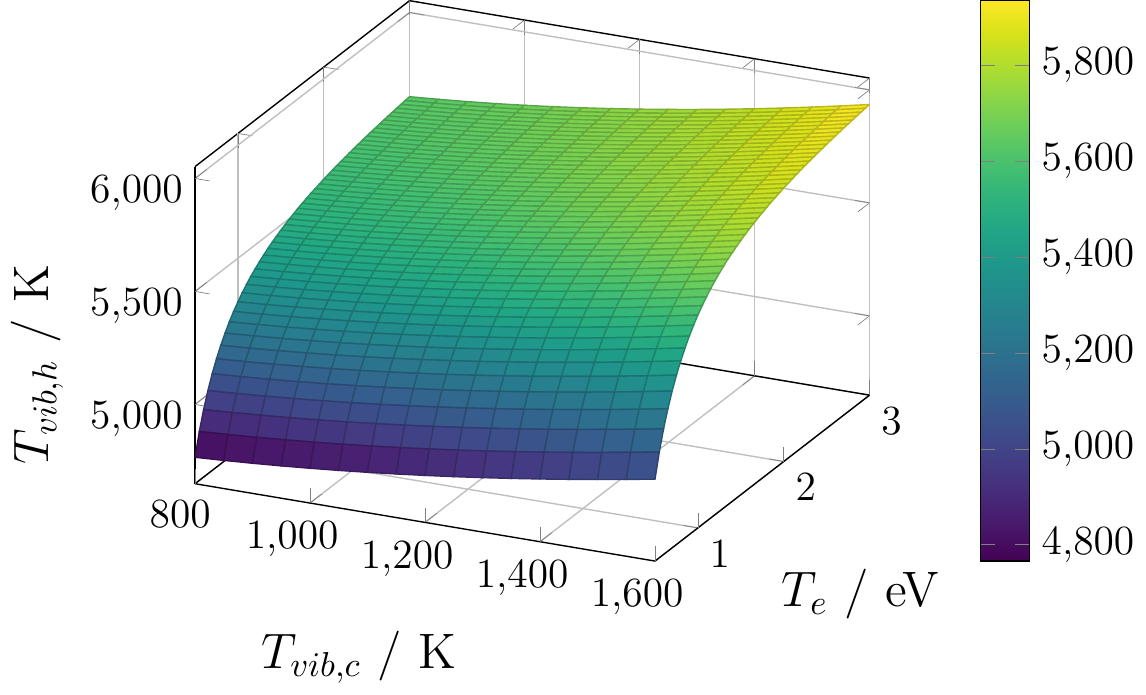}
	\caption{Temperature of the newly excited states in dependence of 
	electron and vibrational temperature before the discharge. }
	\label{fig:Texc_map}
\end{figure}

\begin{figure}
	\centering
	\includegraphics[width=\linewidth]{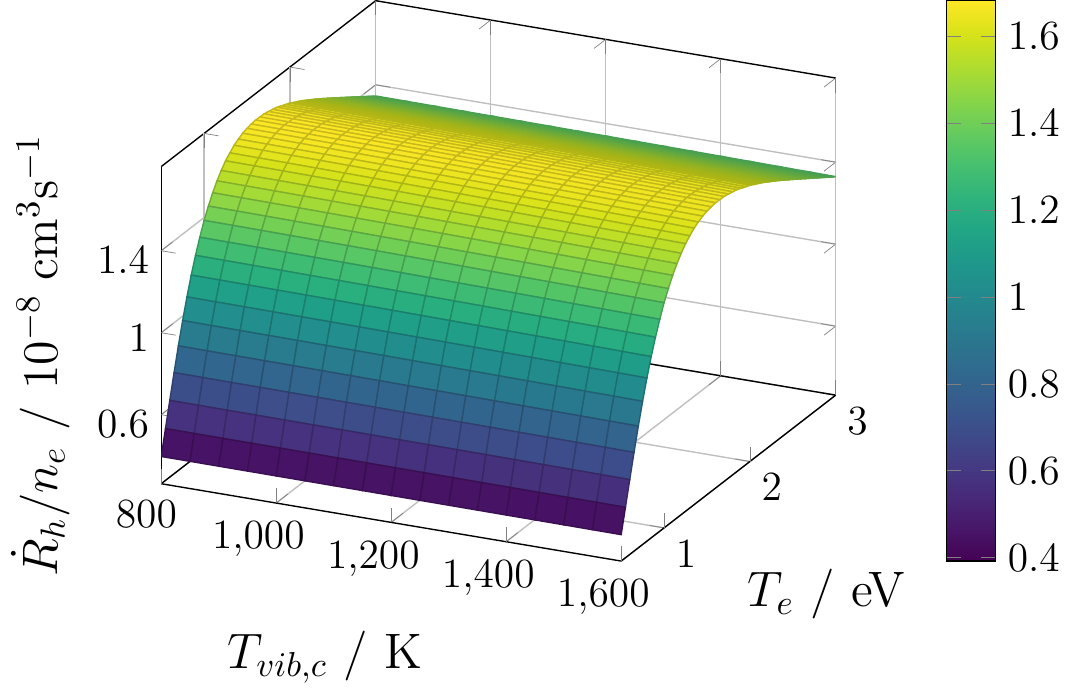}
	\caption{Dependence of $R_h/n_e$ on the temperature of the cold molecules and
	the electrons.}
	\label{fig:Rh_dot_map}
\end{figure}

The two-temperature distribution function in equation \ref{eq:N_twotemp} was already 
introduced in \cite{kuhfeld_vibrational_nodate} but not further motivated. This shall be done here.\\
The main concept of equation \ref{eq:N_twotemp} is that the nitrogen molecules 
can be divided into two mostly independent populations 
\begin{eqnarray}
	N(v,J) = \\ \nonumber
	N_c(v,J;T_{rot}, T_{vib,c}, R_h) + N_h(v,J;T_{rot}, T_{vib,h}, R_h)
\end{eqnarray}
where the population of the vibrationally cold molecules is characterized by 
the cold temperature $T_{vib,c}$ and the population of the hot molecules by the 
hot temperature $T_{vib,h}$. Both populations share the same rotational temperature 
$T_{rot}$ and the fraction of hot and cold molecules compared to the total amount of nitrogen
is given by $R_h$ and $(1-R_h)$ respectively. \\
On time scales of the nanosecond pulse there is essentially no exchange of vibrational
excitation among the nitrogen molecules as V-V and V-T collisions are important 
only on longer time scales. The dominant process of vibrational excitation during
the discharge is most certainly excitation by resonant electron collisions \cite{fridman_plasma_2008}.
This leads to the following interpretation of the two distributions: $N_c$ is the
steady-state background distribution which comprises the bulk of the nitrogen 
molecules. During the discharge some molecules are transferred from 
$N_c$ to $N_h$ by electron collisions. This means the total excitation rate
$\dot{R_h}$ and the vibrational temperature of the newly excited molecules are 
solely defined by the corresponding cross sections, the electron density and the 
electron temperature. \\ 
For the analysis of the measurements in \cite{kuhfeld_vibrational_nodate} the cross section and rate
dataset calculated by Laporta et al 
\cite{laporta_theoretical_2012,laporta_electron-vibration_2013,laporta_electron-impact_2014}
- freely accessible in the Phys4Entry database \cite{noauthor_phys4entry_nodate} - is used.
First the characteristics of the hot distribution are investigated.
While in \eref{eq:N_twotemp} a Boltzmann distribution is assumed, 
there is no obvious physical reason which suggests this choice.
To see that the Boltzmann distribution is still a reasonable good approximation 
for the range of vibrational states visible in this work in the following 
the excitation process is investigated. To begin we assume, that the electron 
conditions during most of the discharge are essentially constant. This is motivated by
the constant electric field in \cite{kuhfeld_vibrational_nodate} and the nearly constant current.
So the rate equation for a vibrational state $v\geq 1$ is given by
\begin{equation}
	\dot{N}_v = \sum_{i<v} N_i k(i,v; T_e) n_e,
	\label{eq:rate_v}
\end{equation}
where $n_e$ is the electron density and $k(i,v;T_e)$ is the rate coefficient 
for the resonant excitation by electron collisions from vibrational state $i$ to $v$ at the electron temperature
$T_e$.
Here, the interest is only in the vibrational dynamics, so particle densities $N_v$ can be understood
as integrated over the rotational quantum number.
Now, we distinguish between the newly excited molecules (i.e. the ones which are
excited during the current discharge pulse) $\tilde N_v$ and the bulk of the molecules  $N_i$
which follows the cold distribution with temperature $T_{vib,c}$.
For constant $n_e$ and $T_e$ the solution of equation \ref{eq:rate_v}
for $\tilde N_v$ with the initial condition $\tilde N_v(t=0) =0$ and under the assumption that the cold
background stays constant, $N_v(t) = \mathrm{const}$, is trivial, and we find that
\begin{eqnarray}
	\tilde N_v &= \left(\sum_{i<v}N_i k(i,v;T_e)\right)n_e t \\\nonumber
	&\propto \sum_{i<v}N_i(T_{vib,c}) k(i,v;T_e)=:\tilde{k}(v;T_{vib,c}, T_e).
	\label{eq:def_kexc}
\end{eqnarray}
This means, that the shape of the distribution
of the hot molecules is constant during the discharge and follows from the values
of the corresponding rates. The sum of rates for all resonant electron processes
into the vibrational states $v$ are depicted in figure \ref{fig:Texc_fit} for an exemplary electron 
temperature of $T_e=\SI{1}{eV}$ and a background temperature of \SI{1300}{K}. 
As can be seen the rates, and therefore populations of the excited states,
follow a similar dependency on the vibrational quantum number
$v$ as a Boltzmann distribution. This motivates the use of a Boltzmann distribution
for the newly excited molecules. In this regard two points should be noted. 
First, the approximation with a Boltzmann distribution is not necessarily usable 
for conditions other than investigated in this work. For states $v>7$ in nitrogen  
significant deviations from a Boltzmann distribution are and also no statements are made 
for other gases here. Second, there is no explicit physical motivation behind the
use of the Boltzmann distribution. Instead, it was chosen because the concept of 
temperatures is convenient and familiar. By more detailed examination of the 
resonant processes leading to vibrational excitation a more precise distribution
might be found which is valid in regimes where this simple approximation 
fails. But for the purpose of this work a Boltzmann distribution is sufficient 
to describe the observed CARS spectra.\\ 
As presented in figure \ref{fig:Texc_fit} the temperature of the hot molecules 
can be obtained by a fit to the rates from \cite{laporta_electron-impact_2014} for
a given vibrational temperature of the molecule bulk and electron temperature.
In figure \ref{fig:Texc_map} values for $T_{vib,h}$ calculated in this way are shown
for a range of $T_{vib,c}$ and $T_e$. As can be seen the dependence on the 
vibrational bulk temperature is weak and decreases further for higher electron 
temperatures.

\begin{figure}
	\centering
	\includegraphics[width=\linewidth]{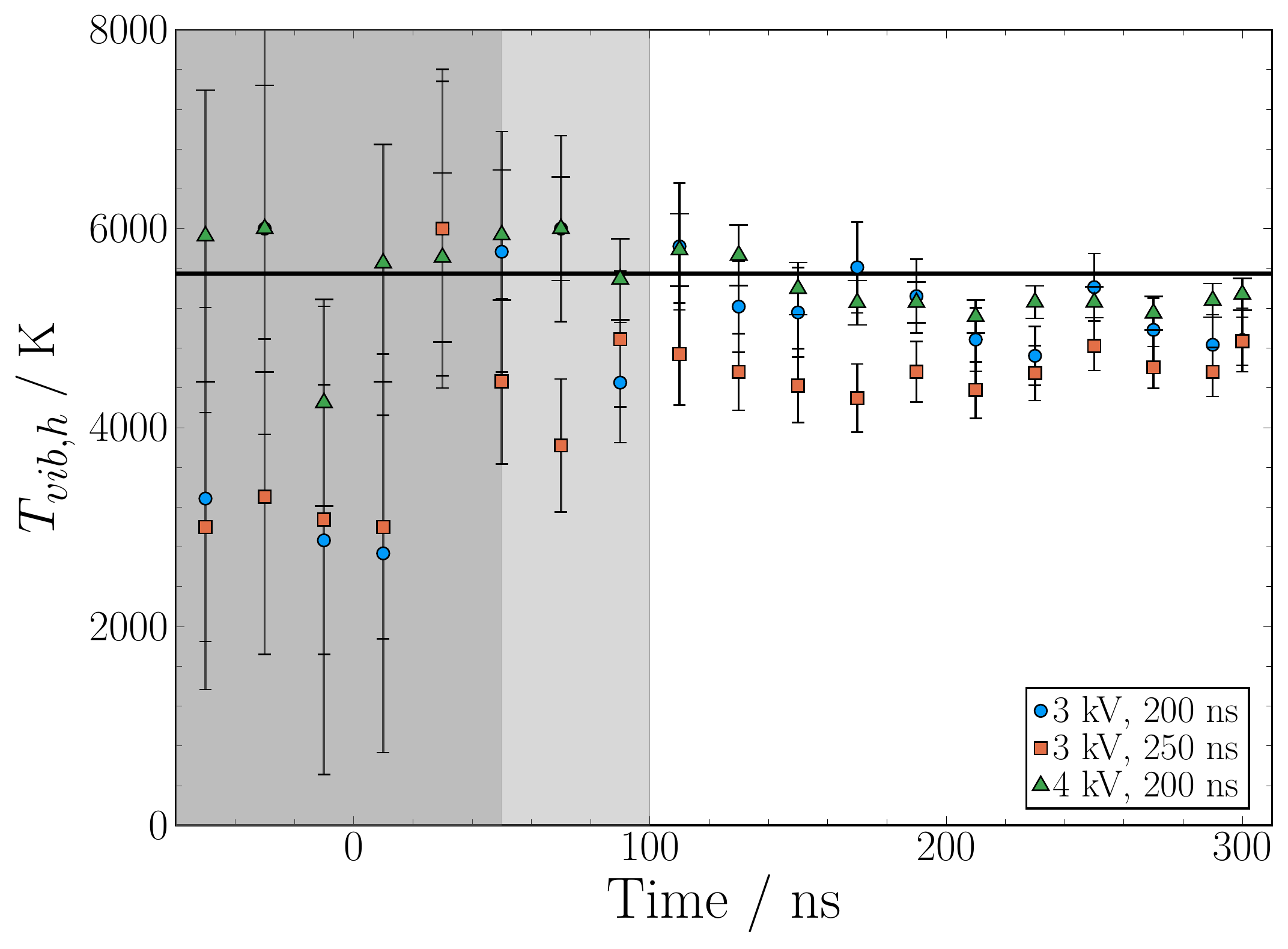}
	\caption{Temperature of the newly excited molecules as obtained by the CARS 
	measurements \cite{kuhfeld_vibrational_nodate}. For the measurement with \SI{4}{kV} and 
	with \SI{3}{kV} applied voltage, the dark gray and the light gray shaded 
	regions should not taken into account, respectively\cite{kuhfeld_vibrational_nodate}. 
	The solid line corresponds to the theoretical
	value for a vibrational bulk temperature of \SI{1300}{K} and a electron temperature
	of \SI{1.2}{eV} (corresponding to a reduced electric field of \SI{81}{Td} 
	according to BOLSIG+ calculations \cite{hagelaar_solving_2005} with the IST Lisbon data set \cite{alves_ist-lisbon_2014}.)}
	\label{fig:Tvh_discharge}
\end{figure}

For a comparison of the theoretical $T_{vib,h}$ the electron temperature is 
estimated with BOLSIG+ \cite{hagelaar_solving_2005} and the IST Lisbon cross section data set \cite{alves_ist-lisbon_2014} to be around
\SI{1.2}{eV} for the measured reduced electric field of \SI{81}{Td}. As the dependency
on $T_{vib,c}$ is weak (see figure \ref{fig:Texc_map}), in figure \ref{fig:Tvh_discharge}
the measurements are only compared to the theoretical value for $T_{vib,c} = \SI{1300}{K}$.
Considering the drastic simplifications in the derivation of the theoretical value
and the measurement uncertainty generally a good agreement can be observed.\\
In addition to the temperature of the newly excited molecules it is possible to 
estimate the rate of excitation. The parameter $R_h$ in \eref{eq:N_twotemp} describes
the total amount of newly excited molecules relative to the gas density.
With \eref{eq:rate_v} one obtains 
\begin{equation}
	\dot{R}_h = \frac{\partial}{\partial t}\left(\sum_{v>0} \tilde N_v \right)= \sum_{v>0}\tilde{k}(v;T_{vib,c}, T_e) n_e
	\label{eq:dot_Rh}
\end{equation}
for the derivative of $R_h$. In figure \ref{fig:Rh_dot_map} the dependence of 
$\dot{R}_h/n_e$ is shown in a similar fashion as it is done for the temperature in 
figure \ref{fig:Texc_map}, and it can be seen that there is essentially no dependence 
on the initial temperature for the excitation rate. 
Strictly speaking, \eref{eq:dot_Rh} is only valid in the beginning of the discharge as
it depends on the approximation $R_h \ll 1$. In our measurements we see an increase 
of $R_h$ up to about 0.1. Therefore, some minor correction need to be applied to 
\eref{eq:dot_Rh}. To account for the depletion of the cold background $N_i$ in 
\eref{eq:def_kexc} is multiplied by $(1-R_h)$. Additionally, when $R_h$ increases
the possibility of multiple vibrational excitation processes can become important.
If molecules from $R_h$ are excited again and end up in states higher than the highest 
state visible in the CARS spectra, $v_{max}$, they cannot be seen in the measurement.
Therefore, an effective loss rate $k_{loss}(v;R_h, T_{vib,h}, T_e)$ is introduced. With these corrections
\eref{eq:dot_Rh} becomes
\begin{eqnarray}
	\dot{R}_h = \sum_{v>0} &\Big[\tilde{k}^\prime(v, R_h;T_{vib,c}, T_e) n_e  \\\nonumber
	& - k_{loss}(v,R_h; T_{vib,h}, T_e)n_e \Big],
	\label{eq:dot_Rh_corrected}
\end{eqnarray}
where $\tilde k^\prime$ and $k_{loss}$ can both be expressed analog to the definition of the 
TTDF \eref{eq:N_twotemp} as
\begin{equation}
	\tilde k^\prime(v,R_h;T_{vib,c}, T_e)=\sum_{i<v}k(i,v;T_e)\frac{1-R_h}{Z_c^\prime}\mathrm{e}^{-\frac{E(i,0)-E(0,0)}{k_BT_{vib,c}}}
\end{equation}
and 
\begin{equation}
	k_{loss}(v,R_h;T_{vib,h}, T_e) = \sum_{j>v_{max}} k(v,j;T_e) \underbrace{\frac{R_h}{Z_h^\prime}\mathrm{e}^{-\frac{E(v,0)-E(0,0)}{k_BT_{vib,h}}}}_{\text{for }v>0}.
\end{equation}
Here, the corresponding distributions are summed over the rotational quantum number which is reflected 
in the partition functions $Z_c^\prime$ and $Z_h^\prime$.
Note, that because $T_{vib,h}(T_{vib,c}, T_e)$, \eref{eq:dot_Rh_corrected} is not more complex than
\eref{eq:dot_Rh} in the sense, that it still depends only on the external parameters $n_e$, $T_{vib,c}$ and 
$T_e$.\\
As the electric field is constant during the current plateau
the electron density can be calculated via 
\begin{equation}
	\frac{I(t)}{A} = \mu(E)E n_e(t).
	\label{eq:I_ne}
\end{equation}
where $I(t)$ is the current measured in the plateau, $E$ the electric field obtained
by the E-FISH measurements\cite{kuhfeld_vibrational_nodate} and $A$ plasma cross section. 
\begin{figure}
	\centering
	\includegraphics[width=\linewidth]{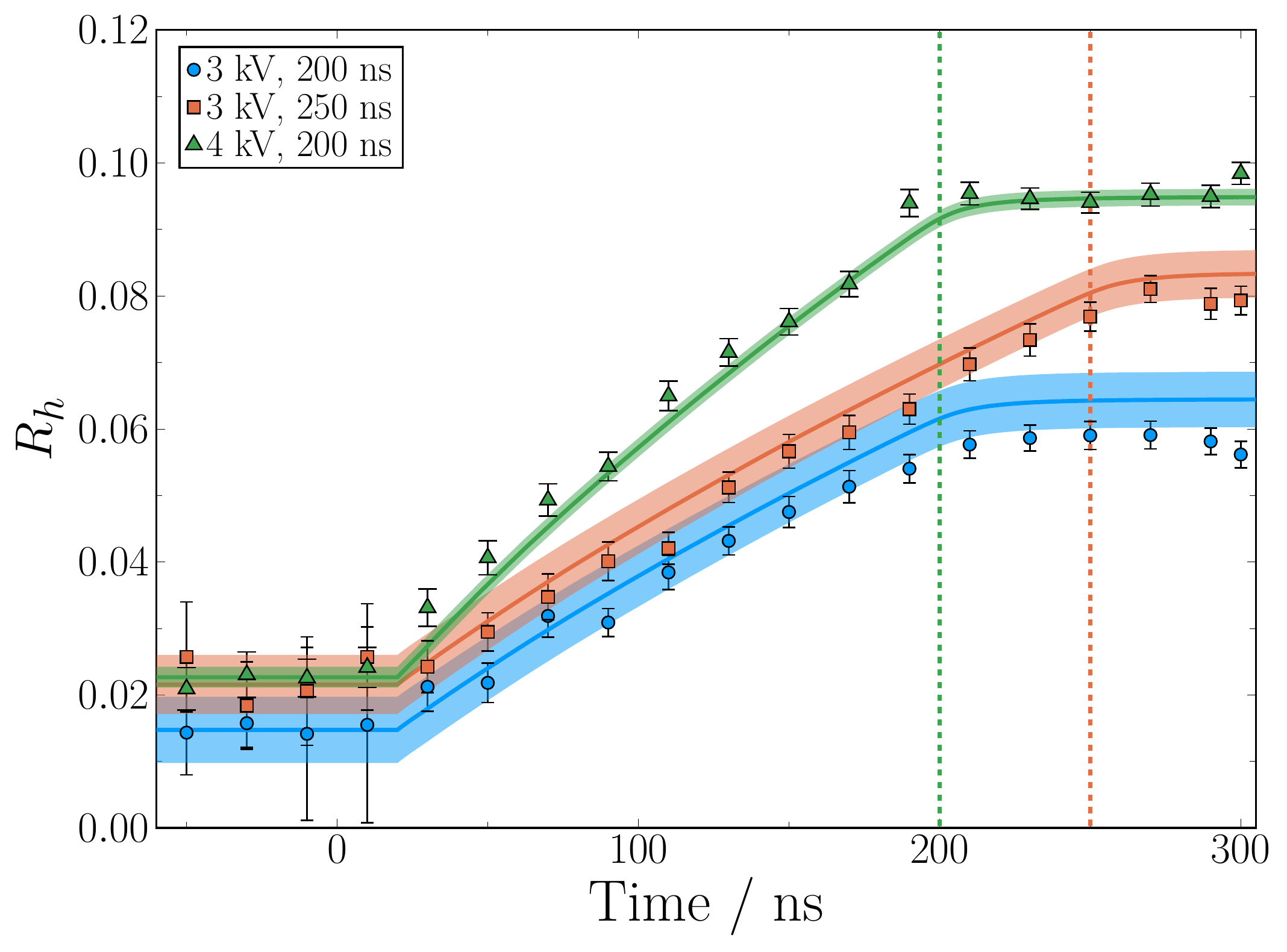}
	\caption{Fraction of the newly excited molecules during the discharge pulse.
	The solid lines are the theoretical rates calculated with the data from 
	\cite{laporta_electron-impact_2014} for electron temperatures and densities 
	obtained by current and field measurements for \SI{4}{kV} applied voltage (green)
	and \SI{3}{kV} (blue). The shaded ribbons give the uncertainty due to the 
	uncertainty in the initial values.}
	\label{fig:Rh_discharge}
\end{figure}
Note, that during the high current pulse in the beginning the electric field can be much 
higher than in the plateau \cite{lepikhin_electric_2020}. Therefore, \eref{eq:I_ne} is 
only evaluated for $t>\SI{20}{ns}$, i.e. during the plateau phase.
In figure \ref{fig:Rh_discharge} $R_h$ calculated by integrating \eref{eq:dot_Rh_corrected} is compared to the 
measured values of $R_h$ during the discharge pulse. The initial value 
for the integration is chosen to be the average of the three data points 
before the ignition of the discharge ($t<\SI{0}{ns}$). A very good agreement is 
observed between the measurements and the theoretical values calculated from 
the current and field measurements even though those are completely independent 
of the CARS measurements. 

\subsection{Afterglow}
\label{sec:afterglow}

To analyze the dynamics of vibrationally excited states in the afterglow between
two discharge pulses a kinetic model is developed for vibrationally excited nitrogen
up to $v=9$.\\
As the ionization degree is very low, superelastic collisions with the plasma electrons
are neglected here. Furthermore, the influence of electronically excited molecules
is ignored and the dissociation degree is assumed to be small. These assumptions
reduce the reaction set to V-V and V-T collisions among the nitrogen molecules.\\
The rates for the V-V process
\begin{equation}
	\textrm{N}_2(v+1) + \textrm{N}_2(w) 
	\rightarrow \textrm{N}_2(v) +\textrm{N}_2(w+1)
\end{equation}
are calculated from the rate for the process
\begin{equation}
	\textrm{N}_2(1) + \textrm{N}_2(0) 
	\rightarrow \textrm{N}_2(0) +\textrm{N}_2(1)
\end{equation}
via the scaling law of the semiclassical forced harmonic oscillator 
(FHO) model\cite{adamovich_three-dimensional_2001,ahn_determination_2004}
\begin{eqnarray}
	k_{VV}(v+1, w \rightarrow v, w+1) = \\\nonumber
	k_{VV}(1,0 \rightarrow 0,1) (v+1)(w+1) \frac{3-\textrm{e}^{-\frac{2\lambda}{3}}}{2}
	\textrm{e}^{-\frac{2\lambda}{3}}\exp\left(\frac{\Delta E}{2k_B T}\right)
\end{eqnarray}
where $k_B$ is the Boltzmann constant and $T$ the gas temperature.
$\lambda$ is given by \cite{ahn_determination_2004}
\begin{equation}
	\lambda = \frac{1}{3\sqrt{2}} \sqrt{\frac{\theta}{T}}\frac{|\Delta E|}{\omega\hbar}
\end{equation}
with the harmonic angular frequency $\omega$ and \mbox{$\theta=\frac{4\pi^2\omega^2 m}{\alpha^2k_B}$}. 
$\Delta E = E(v+1) + E(w) - E(v) - E(w+1)$ is the energy defect due to the 
anharmonicity of the vibrational potential, $m$ is the collisional reduced mass and
$\alpha$ is the exponential repulsive potential parameter \cite{ahn_determination_2004}.
For $k_{VV}(1,0\rightarrow 0,1)$ the value \SI{0.9e-14}{\cubic\centi\metre\per\second} is used \cite{billing_vv_1979}.\\
For completeness the V-T rates from the calculations provided by Billing and Fisher \cite{billing_vv_1979} for a 
(rotational-translational) gas temperature of \SI{300}{K} are included.
It should be noted that they are several orders of magnitude smaller and do not 
affect the simulation in a noticeable way.\\
The simulation volume is one half of the discharge jet, i.e. a cuboid with edge lengths
$\SI{1}{mm}\times\SI{1}{mm}\times\SI{10}{mm}$. As first approximation the particle
densities are assumed to be homogeneous in the whole volume.
The influx of is considered via \cite{kemaneci_global_2014}
\begin{equation}
	\left.\frac{dN_0}{dt}\right|_{in} = N_{gas}\frac{CQP_{atm}}{VP}
	\label{eq:influx}
\end{equation}
where $Q=\SI{10}{sccm}$ is half of the set gas flow - as it is assumed that the
total gas flow of \SI{20}{sccm} is split equally to both sides of the jet. 
$C=\SI{1.6667e-8}{\cubic\metre\per\second\per sccm}$ is the conversion factor to 
convert \si{sccm} to \si{\cubic\metre\per\second}, $P_{atm}=\SI{1}{bar}$ is the 
atmospheric pressure and $P = \SI{200}{mbar}$ is the pressure in the discharge chamber 
and $N_{gas}$ is the gas density.
In \eref{eq:influx} it is assumed, that the inflowing particles are all in the 
vibrational ground state as at room temperature there is now significant vibrational
excitation.
The vibrational excited nitrogen molecules in state $v$ exiting the jet are described
by
\begin{equation}
	\left.\frac{dN_v}{dt}\right|_{out} = -N_{v}\frac{CQP_{atm}}{VP}.
	\label{eq:outflux}
\end{equation}
The loss of excited molecules by diffusion to the walls is given by \cite{chantry_simple_1987,kemaneci_global_2014}
\begin{eqnarray}
	\left.\frac{\partial N_{v>0}}{\partial t}\right|_{W} &= -N_{v>0}
	\left(\frac{\Lambda_0^2}{D_v} + \frac{2(2-\gamma_v)V}{v_{th}\gamma_v A}\right)^{-1} \\\nonumber
	&\approx -\frac{v_{th}\gamma_v A}{2(2-\gamma_v)V} 
	\label{eq:wall}
\end{eqnarray}
with the characteristic length scale for the given geometry $\Lambda_0$ \cite{chantry_simple_1987}, 
the volume and surface area $V$ and $A$, the diffusion
coefficient for the species $D_v$ and the corresponding deactivation coefficient $\gamma_v$
giving the probability to lose a particle in state $v$ when it hits the wall.
$\gamma_v$ is not very well known for states $v>1$ but for $v=1$ it is typically
in the order of about \SI{4e-4}{} to \SI{3e-3}{} for different materials \cite{black_measurements_1974}.
In absence of better knowledge the deactivation coefficient is chosen here to 
be $\gamma=\SI{1e-3}{}$ for all states. This motivates the approximation in 
\eref{eq:wall} which is consistent with the assumption of flat density
profiles: the very low deactivation coefficient means that the loss of excited 
particles is not limited by the diffusion, but instead by the deactivation
process once the molecules reach the walls. Finally, we assume that the deactivation
happens directly into the ground state, so that the walls are effectively a source
for $v=0$:
\begin{equation}
	\left.\frac{\partial N_0}{\partial t}\right|_{W} = 
	-\sum_{v>0}\left.\frac{\partial N_v}{\partial t}\right|_{W}
	\label{eq:wall0}
\end{equation}

\begin{figure}
	\centering
	\includegraphics[width=\linewidth]{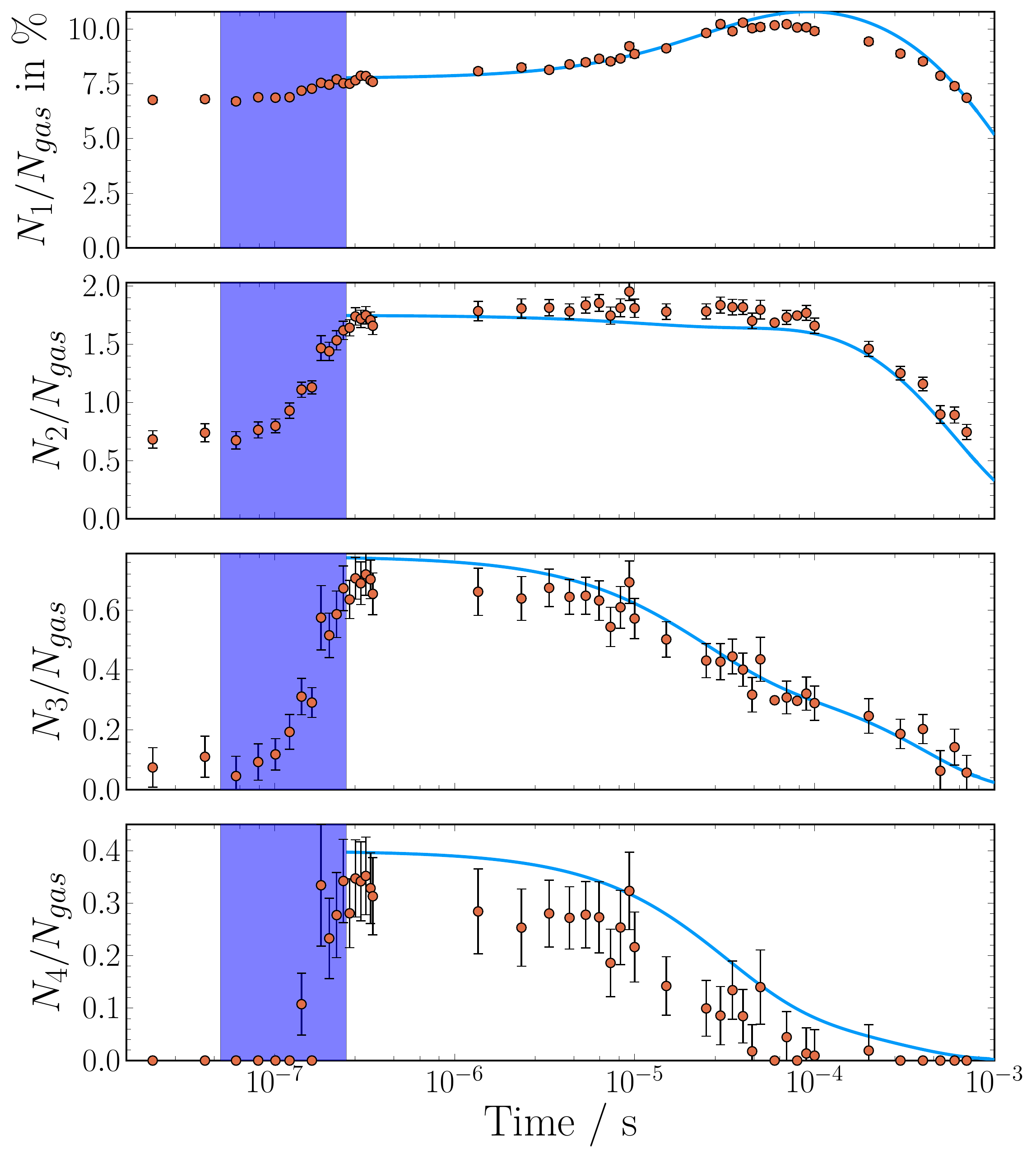}
	\caption{Relative population densities for the measurement \SI{3}{kV}, \SI{200}{ns}.
	Symbols are the population densities obtained with the method described in \cite{kuhfeld_vibrational_nodate}
	and lines are the results from the kinetic model.}
	\label{fig:afterglow_2}
\end{figure}

The initial conditions for the simulation are obtained from the two temperature 
distribution directly after the discharge pulse. In this way the initial population densities
are extrapolated up to $v=9$.
The results of the simulation are compared to the measurement results from \cite{kuhfeld_vibrational_nodate}
in figure \ref{fig:afterglow_2}. For clarity of presentation the results for the 
other two measurement conditions are shown in the appendix (see figures \ref{fig:afterglow_3}
and \ref{fig:afterglow_4}). Those show a similar good agreement.

From the very good agreement between the measurements and the simple simulation
it can be inferred, that here the E-V transitions, i.e. population of vibrational 
states by deexcitation of electronically excited molecules,
seem to play only a minor role in contrast to previous works \cite{deviatov_investigation_1986, montello_picosecond_2013}. 
A possible reason to explain this difference could be a generally higher electric field 
in discharges therein, which increases the amount of electronic excitation. 
While there is not much information
about the discharge used by \mbox{Deviatov \etal \cite{deviatov_investigation_1986},} 
\mbox{Montello \etal \cite{montello_picosecond_2013}} used a pin-to-pin 
discharge with a \SI{10}{mm} gap. Their estimated electric field reaches up to 
about \SI{275}{Td} during the ignition phase and stays at about \SI{125}{Td} during
the discharge plateau which is significantly higher than the value obtained by E-FISH measurements
\cite{kuhfeld_vibrational_nodate} in the discharge investigated here. 
Additionally, their pulse is only \SI{150}{ns} long which 
further increases the relative importance of the high electric field during the ignition.
Therefore, it is very likely, that in their discharges the density of electronically excited
states relative to the vibrationally excited states is significantly higher.

\section{Conclusion}
In this paper, the vibrational two-temperature distribution function used in the companion
paper \cite{kuhfeld_vibrational_nodate} to evaluate CARS spectra during a ns-pulsed discharge 
was motivated and the different parameters were connected 
to the underlying physical processes. It is found that under the investigated 
discharge conditions the direct resonant excitation through electron collisions is 
the main path for production of vibrationally excited nitrogen. The distribution 
function of the newly excited molecules follows therefore the shape of the corresponding
excitation probabilities or rates. In the case of resonant excitation this shape
closely resembles a Boltzmann distribution for small $v$, leading to the two-temperature distribution
function \eref{eq:N_twotemp} consisting of a Boltzmann distributed cold background
and the also Boltzmann distributed newly excited hot molecules. For the case that the excitation
rates do not follow a Boltzmann distribution the hot part of the distribution 
function can be modified easily. If sufficient data for the excitation process is 
known the parameter in the distribution function can be estimated, and it is found
that the estimates - using the rates reported by Laporta \cite{laporta_electron-impact_2014} in 
combination with current and field measurements - agree very well with the measured
values. This shows that the rather simple two-temperature distribution - 
and its generalization by allowing non-Boltzmann distributions - provides a useful
framework for analyzing and potentially optimizing the vibrational excitation of 
molecules in plasmas where the resonant excitation by electron collisions is the 
dominant process.
Furthermore, the fact that the current and electric field are nearly constant
during almost the whole high voltage pulse 
for the discharge type investigated in this work, indicates a constant
electron density during the majority of the discharge which is created essentially 
only during the ignition of the pulse. This promises an easy tool for estimating 
the amount of vibrational excitation a priori when one is able to predict the 
density and electric field for the given discharge. \\
For the afterglow it was found, that a relative simple model considering only
V-V and V-T transfer and transport losses is enough to reproduce the measurements.
Meaning E-V transfer - where vibrational excited molecules in the electronic ground
state are produced by cascades or quenching of higher electronic states - seems to 
be of minor importance in this discharge type. A possible reason why E-V transfer 
was needed to explain previous vibrational measurements \cite{deviatov_investigation_1986, montello_picosecond_2013} could be the generally
higher electric field in those works. \\
Finally, it can be concluded that the discharge reactor investigated here
shows promise to be a handy tool while investigating the influence of vibrational 
excitation for example on plasma chemistry or plasma catalysis. Together with 
the description of the vibrational system provided here it allows to easily control
and understand the vibrational excitation in the plasma volume, which is vital to 
understand its influence on other reactions.

\section*{Acknowledgements}
This project is supported by the DFG (German Science Foundation) within the 
framework of the CRC (Collaborative Research Centre) 1316 "Transient atmospheric
plasmas - from plasmas to liquids to solids". 

\appendix
\section{Additional simulation results}
\begin{figure}[h]
	\centering
	\includegraphics[width=\linewidth]{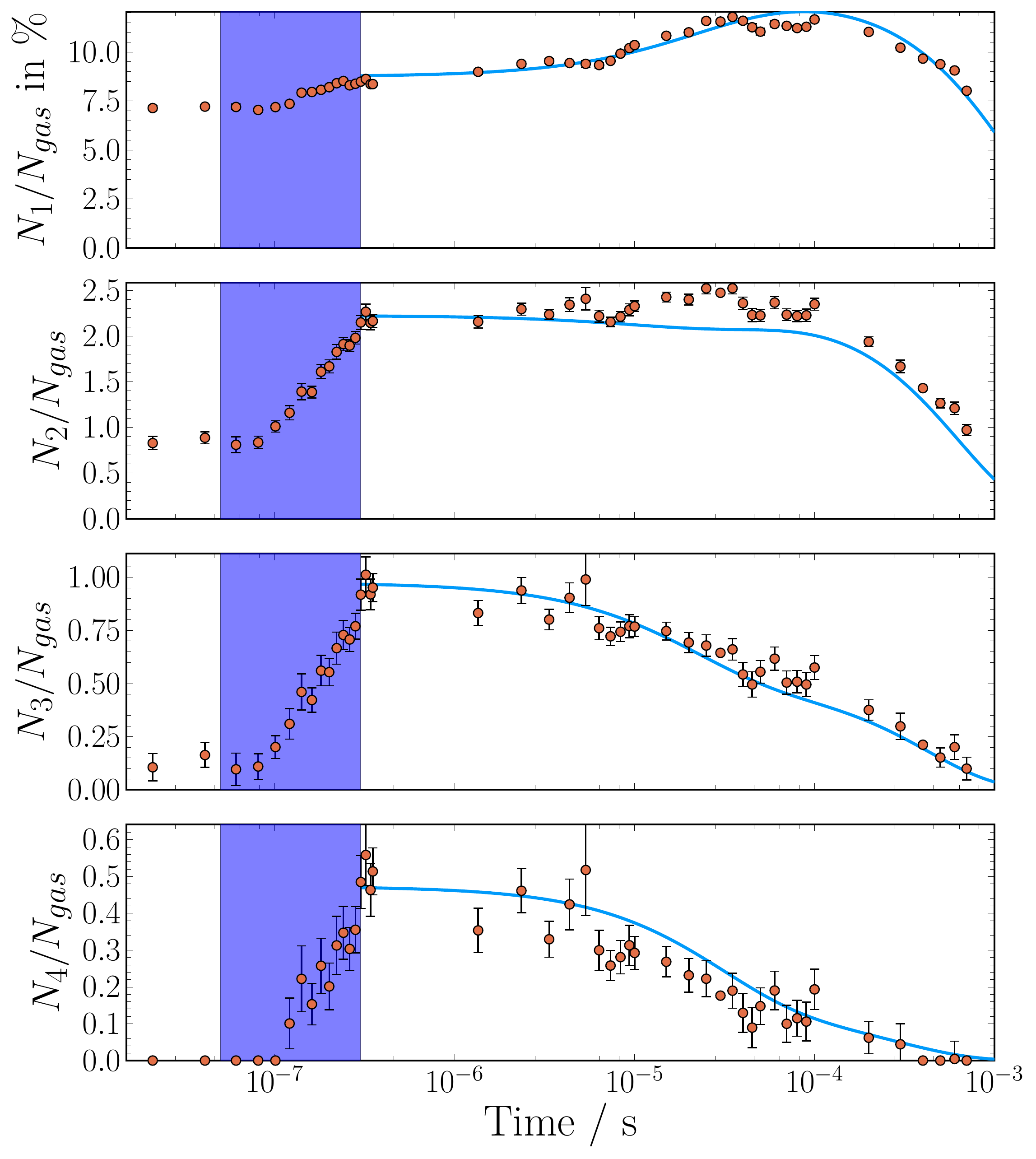}
	\caption{Relative population densities for the measurement \SI{3}{kV}, \SI{250}{ns}.
	Symbols are the population densities obtained with the method described in \cite{kuhfeld_vibrational_nodate}
	and lines are the results from the kinetic model.}
	\label{fig:afterglow_3}
\end{figure}

\begin{figure}[h]
	\centering
	\includegraphics[width=\linewidth]{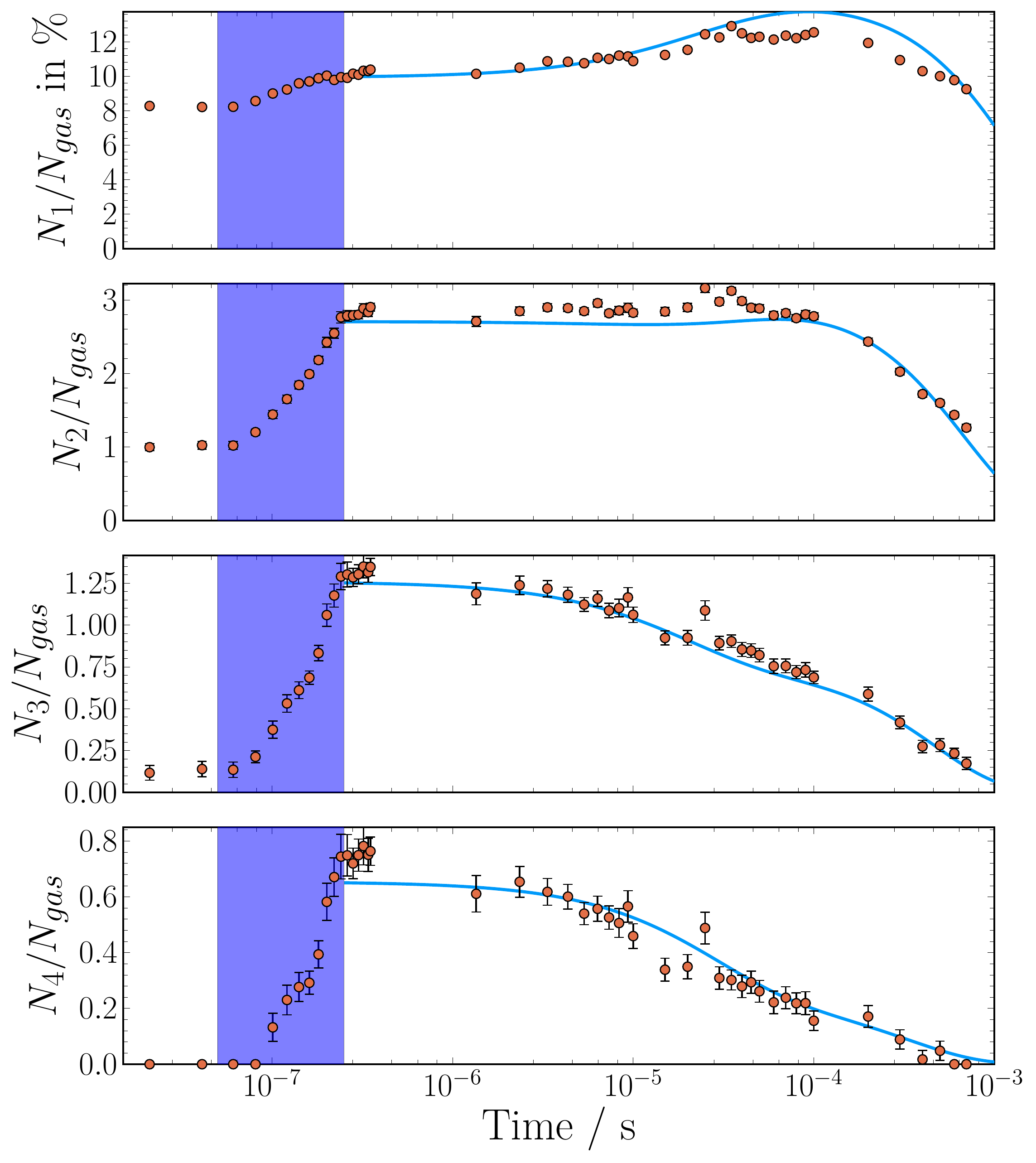}
	\caption{Relative population densities for the measurement \SI{4}{kV}, \SI{200}{ns}.
	Symbols are the population densities obtained with the method described in \cite{kuhfeld_vibrational_nodate}
	and lines are the results from the kinetic model.}
	\label{fig:afterglow_4}
\end{figure}

For completeness here the results of the simulation are compared to the measurements "\SI{3}{kV}, \SI{200}{ns}" and
"\SI{4}{kV}, \SI{200}{ns}" from \cite{kuhfeld_vibrational_nodate} in figures~\ref{fig:afterglow_3} and
\ref{fig:afterglow_4}. The agreement is as good as in figure
\ref{fig:afterglow_2}, leading to the conclusion, that for all conditions investigated the 
processes included in the simulation are sufficient to explain the experimental results. 


\section*{References}
\bibliography{references}

\providecommand{\newblock}{}
\begin{thebibliography}{10}
\expandafter\ifx\csname url\endcsname\relax
  \def\url#1{{\tt #1}}\fi
\expandafter\ifx\csname urlprefix\endcsname\relax\def\urlprefix{URL }\fi
\providecommand{\eprint}[2][]{\url{#2}}

\bibitem{kuhfeld_vibrational_nodate}
Kuhfeld J, Lepikhin N~D, Luggenhölscher D and Czarnetzki U Vibrational {CARS}
  measurements in a near-atmospheric pressure plasma jet in nitrogen: {I}.
  {Measurement} procedure and results

\bibitem{laporta_electron-impact_2014}
Laporta V, Little D~A, Celiberto R and Tennyson J 2014 Electron-impact resonant
  vibrational excitation and dissociation processes involving vibrationally
  excited {N2} molecules {\em Plasma Sources Science and Technology\/} {\bf 23}
  065002

\bibitem{lepikhin_electric_2020}
Lepikhin N~D, Luggenhölscher D and Czarnetzki U 2020 Electric field
  measurements in a {He}:{N2} nanosecond pulsed discharge with sub-ns time
  resolution {\em Journal of Physics D: Applied Physics\/} {\bf 54} 055201

\bibitem{hagelaar_solving_2005}
Hagelaar G~J~M and Pitchford L~C 2005 Solving the {Boltzmann} equation to
  obtain electron transport coefficients and rate coefficients for fluid models
  {\em Plasma Sources Science and Technology\/} {\bf 14} 722--733

\bibitem{fridman_plasma_2008}
Fridman A~A 2008 {\em Plasma chemistry\/} first paperback edition ed
  (Cambridge: Cambridge University Press)

\bibitem{neyts_understanding_2014}
Neyts E~C and Bogaerts A 2014 Understanding plasma catalysis through modelling
  and simulation—a review {\em Journal of Physics D: Applied Physics\/} {\bf
  47} 224010

\bibitem{whitehead_plasmacatalysis_2016}
Whitehead J~C 2016 Plasma–catalysis: the known knowns, the known unknowns and
  the unknown unknowns {\em Journal of Physics D: Applied Physics\/} {\bf 49}
  243001

\bibitem{lempert_coherent_2014}
Lempert W~R and Adamovich I~V 2014 Coherent anti-{Stokes} {Raman} scattering
  and spontaneous {Raman} scattering diagnostics of nonequilibrium plasmas and
  flows {\em Journal of Physics D: Applied Physics\/} {\bf 47} 433001

\bibitem{eckbreth_laser_1996}
Eckbreth A~C 1996 {\em Laser diagnostics for combustion temperature and
  species\/} 2nd ed ({\em Combustion science and technology book series\/} no
  volume 3) (Boca Raton: CRC Press)

\bibitem{shaub_direct_1977}
Shaub W~M, Nibler J~W and Harvey A~B 1977 Direct determination of
  non‐{Boltzmann} vibrational level populations in electric discharges by
  {CARS} {\em The Journal of Chemical Physics\/} {\bf 67} 1883--1886

\bibitem{deviatov_investigation_1986}
Deviatov A~A, Dolenko S~A, Rakhimov A~T, Rakhimova T~V and Roi N~N 1986
  Investigation of kinetic processes in molecular nitrogen by the {CARS}
  technique {\em Zhurnal Eksperimentalnoi i Teoreticheskoi Fiziki\/} {\bf 90}
  429--436

\bibitem{montello_picosecond_2013}
Montello A, Yin Z, Burnette D, Adamovich I~V and Lempert W~R 2013 Picosecond
  {CARS} measurements of nitrogen vibrational loading and
  rotational/translational temperature in non-equilibrium discharges {\em
  Journal of Physics D: Applied Physics\/} {\bf 46} 464002

\bibitem{messina_study_2007}
Messina D, Attal-Trétout B and Grisch F 2007 Study of a non-equilibrium pulsed
  nanosecond discharge at atmospheric pressure using coherent anti-{Stokes}
  {Raman} scattering {\em Proceedings of the Combustion Institute\/} {\bf 31}
  825--832

\bibitem{laporta_theoretical_2012}
Laporta V, Celiberto R and Wadehra J~M 2012 Theoretical vibrational-excitation
  cross sections and rate coefficients for electron-impact resonant collisions
  involving rovibrationally excited {N2and} {NO} molecules {\em Plasma Sources
  Science and Technology\/} {\bf 21} 055018

\bibitem{laporta_electron-vibration_2013}
Laporta V and Bruno D 2013 Electron-vibration energy exchange models in
  nitrogen-containing plasma flows {\em The Journal of Chemical Physics\/} {\bf
  138} 104319

\bibitem{noauthor_phys4entry_nodate}
{PHYS4ENTRY} (7th {Framework} {Programme})
  \urlprefix\url{https://users.ba.cnr.it/imip/cscpal38/phys4entry/database.html}

\bibitem{alves_ist-lisbon_2014}
Alves L~L 2014 The {IST}-{LISBON} database on {LXCat} {\em Journal of Physics:
  Conference Series\/} {\bf 565} 012007

\bibitem{adamovich_three-dimensional_2001}
Adamovich I~V 2001 Three-{Dimensional} {Analytic} {Model} of {Vibrational}
  {Energy} {Transfer} in {Molecule}-{Molecule} {Collisions} {\em AIAA
  Journal\/} {\bf 39} 1916--1925

\bibitem{ahn_determination_2004}
Ahn T, Adamovich I~V and Lempert W~R 2004 Determination of nitrogen {V}–{V}
  transfer rates by stimulated {Raman} pumping {\em Chemical Physics\/} {\bf
  298} 233--240

\bibitem{billing_vv_1979}
Billing G~D and Fisher E~R 1979 {VV} and {VT} rate coefficients in {N2} by a
  quantum-classical model {\em Chemical Physics\/} {\bf 43} 395--401

\bibitem{kemaneci_global_2014}
Kemaneci E, Carbone E, Booth J~P, Graef W, Dijk J~v and Kroesen G 2014 Global
  (volume-averaged) model of inductively coupled chlorine plasma: {Influence}
  of {Cl} wall recombination and external heating on continuous and
  pulse-modulated plasmas {\em Plasma Sources Science and Technology\/} {\bf
  23} 045002

\bibitem{chantry_simple_1987}
Chantry P~J 1987 A simple formula for diffusion calculations involving wall
  reflection and low density {\em Journal of Applied Physics\/} {\bf 62}
  1141--1148

\bibitem{black_measurements_1974}
Black G, Wise H, Schechter S and Sharpless R~L 1974 Measurements of
  vibrationally excited molecules by {Raman} scattering. {II}. {Surface}
  deactivation of vibrationally excited {N2} {\em The Journal of Chemical
  Physics\/} {\bf 60} 3526--3536

\end{thebibliography}

\end{document}